\documentclass[sigconf]{acmart}

\usepackage{booktabs} 
\usepackage{subcaption,graphicx}
\usepackage{appendix,subcaption}

\setcopyright{acmcopyright}
\usepackage{todonotes}
\usepackage{url}

\acmDOI{arXiv:1703.06324}

\acmConference[KDD Workshop on ML meets Fashion]{KDD}{August 2017}{Canada} 
\acmYear{2017}
\copyrightyear{2017}

\begin{document}
\title{Deep Tensor Encodings}

\author{B Sengupta}
\authornote{BS has dual appointments at Cortexica Vision Systems Ltd. and the Dept. of Bioengineering, Imperial College London.}
\affiliation{%
  \institution{Dept. of Engineering, University of Cambridge}
  \streetaddress{Trumpington Street}
 \city{Cambridge} 
  \state{United Kingdom}
  \postcode{CB2 1PZ}
}
\email{bs573@cam.ac.uk}

\author{Y Qian}
\authornote{BS and YQ contributed equally to this paper.}
\author{E Vazquez}
\affiliation{%
  \institution{Cortexica Vision Systems Limited}
  \streetaddress{Capital Tower -- 91 Waterloo Road}
  \city{London} 
  \state{United Kingdom} 
  \postcode{SE1 8RT}
}
\email{yu.qian@cortexica.com}

\renewcommand{\shortauthors}{Sengupta \textit{et al.}}

\begin{abstract} 
Learning an encoding of feature vectors in terms of an over-complete dictionary or a  information geometric (Fisher vectors) construct is wide-spread in statistical signal processing and computer vision. In content based information retrieval using deep-learning classifiers, such encodings are learnt on the flattened last layer, without adherence to the multi-linear structure of the underlying feature tensor.  We illustrate a variety of feature encodings incl. sparse dictionary coding and Fisher vectors along with proposing that a structured tensor factorization scheme enables us to perform retrieval that can be at par, in terms of average precision, with  Fisher vector encoded image signatures. In short, we illustrate how structural constraints increase retrieval fidelity.
\end{abstract} 

%
%




\maketitle
\begin{abstract}
Learning an encoding of feature vectors in terms of an over-complete dictionary or an information geometric (Fisher vectors) construct is wide-spread in statistical signal processing and computer vision. In content based information retrieval using deep-learning classifiers, such encodings are learnt on the flattened last layer, without adherence to the multi-linear structure of the underlying feature tensor.  We illustrate a variety of feature encodings incl. sparse dictionary coding and Fisher vectors along with proposing that a structured tensor factorization scheme enables us to perform top item retrieval that is at par, in terms of average precision, with  Fisher vector encoded image signatures. 
\end{abstract}

\section{Introduction}

The success of deep-learning lies in constructing feature spaces where in competing classes of objects, sounds, etc. can be shattered using high-dimensional hyperplanes. The classifier relies on the accumulation of representation in the final convolution layer of a deep neural network. Often times, the classifier performance increases as one incorporates information from earlier layers of the neural network. Such a structural constraint has been imposed on certain deep learning architectures via the \textit{inception module}. In addition to decreasing the computational effort, utilization of $1 \times 1$ convolution filters enables the dimensionality of feature map to be immensely reduced; in tandem with pooling, the dimensionality reduces even further. Thus, information from the previous layer(s) can be accumulated and concatenated in the inception module. By learning the weights feeding to the inception module there is the additional flexibility in vetting the different sources of information that reaches deeper layers. 

A big demerit of deep-learning architectures is their inability to perform well in the advent of small amounts of training data. Tricks such as input (rotation, blur, etc.) and feature augmentation (in terms of inception module) have proven useful \cite{Goodfellow2016}. Such structural constraints regularize the optimization problem, reducing over-fitting. In this paper, we propose a much simpler structural constraint i.e., to utilize the multi-linear structure of deep feature tensors. We will first emphasize the importance of feature encoding, starting with Fisher encoding (of the last layer) followed by a sparse dictionary based on the last feature layer; this feeds to deep tensor factorization that brings together tensor learning and deep learning -- cementing the idea that taking into account the high dimensional multi-linear representation increases the fidelity of our learnt representation.  Albeit, these algorithms are evaluated on a content-based-image-retrieval (CBIR) setting for a texture dataset, many of them have been combined in Cortexica's findSimilar technology (\url{https://www.cortexica.com/technology/}; Figures \ref{fig:fsimilar} and \ref{fig:simexamps}), that facilitate retailers to recommend items from fashion databases comprising inventory items such as tops, trousers, handbags, etc.  

\section{Methods}

\subsection{Dataset and deep-feature generation}
\label{sec:dataset}

In this paper, Describable Textures Dataset (DTD) \cite{Cimpoi2014} is used to evaluate feature encodings for image retrieval. A wide variety of fashion inventory rely on capturing the differences between varying textures. Hence, our feature encoding comparison leverages the DTD dataset, a widely used dataset for texture discrimination. Rather than recognition and description of object, texture images in DTD are collected from wild images (Google and Flickr) and classified based on human visual perception \cite{Tamura1978}, such as directionality (line-like), regularity (polka-dotted and chequered), etc.  DTD contains 5640 wild texture images with 47 describable attributes drawn from the psychology literature, and is publicly available on the web at \url{http://www.robots.ox.ac.uk/vgg/data/dtd/}. 

Textures can be described via orderless pooling of filter bank response \cite{Gong2014}. In deep convolutional neural network (dCNN), the convolutional layers are akin to non-linear filter banks; these have in fact been proved to be better for texture descriptions \cite{Cimpoi2015}. Here, the local deep features are extracted from last convolutional layer of a pre-trained VGG-M \cite{Chatfield2014}. This is represented by $T = \left ( t_{1},...t_{i},...,t_{N}: t \in \mathbb{R}^{D} \right )$; the size of last convolutional layer is $H\times W\times D$, where $D$ denotes the dimension of filter response at the $i^{th}$ pixel of last convolution layer; $N = H\times W$ is the total number of local features. Different feature encoding strategies are then utilized for encoding local descriptors. A  similarity metric is then applied to rank images. We use the $l_2$ norm (Frobenius norm for matrices and tensors) between vectors as a notion of distance between the query and the database images. 

We will now introduce five different encodings of the feature matrix -- (a) Fisher encoding, (b) Sparse matrix dictionary learning, (c) t-SVD factorization, (d) Low-rank plus Sparse factorization, and e) Multilinear Principal Component Analysis (MPCA). 

\subsection*{Feature encoding}
\label{sec:feature_encoding}

\subsubsection{Fisher encoding}
\label{sec:fisher_encoding}

We use a Gaussian Mixture Model (GMM) for encoding a probabilistic visual vocabulary on the training dataset. Images are then represented as Fisher Vectors \cite{Perronnin2006,Uricchio2015} -- derivatives of log-likelihood of the model with respect to its parameters (the score function). Fisher encoding describes how the distribution of features of an individual image differs from the distribution fitted to the feature of all training images.
 
First, a set of $D$ dimension local deep features are  extracted from an image and denoted as $T = \left ( t_{1},...t_{i},...,t_{N}: t \in \mathbb{R}^{D} \right )$. As Ref. \cite{Simonyan13,Cimpoi2015}, a $K$ component GMM  with diagonal covariance is generated on the training set with the parameters $\left \{ \Theta = \left (\omega _{k}, \mu _{k},\Sigma _{k} \right ) \right \}_{k=1}^{K}$, only the derivatives with respect to the  mean $\left \{ \mu _{k} \right \}_{k=1}^{K}$ and variances $\left \{ \Sigma _{k} \right \}_{k=1}^{K}$ are encoded and concatenated to represent an image $T\left ( X,\Theta  \right ) = \left ( \frac{\partial L}{\partial \mu_{1}},...,\frac{\partial L}{\partial \mu_{K}},\frac{\partial L}{\partial \Sigma _{1}},...,\frac{\partial L}{\partial \Sigma _{K}} \right )$, where,
 
\begin{eqnarray}
L\left ( \Theta  \right ) & = & \sum _{i=1}^{N} log\left ( \pi\left ( t_{i} \right ) \right ) \nonumber \\
\pi\left ( t_{i} \right ) & = & \sum_{k=1}^{K}\omega _{k}N\left ( t_{i};\mu _{k} ,\Sigma _{k}\right )
\end{eqnarray}

For each component $k$, mean and covariance deviation on each vector dimension $j=1,2...D$ are 

\begin{eqnarray}
\frac{\partial L}{\partial \mu_{jk}} & = & \frac{1}{N\sqrt{\omega _{k}}}\sum _{i=1}^{N}q_{ik}\frac{t_{ji}-\mu _{jk}}{\sigma_{jk}  } \nonumber \\
\frac{\partial L}{\partial \Sigma _{jk}} & = & \frac{1}{N\sqrt{2\omega _{k}}}\sum _{i=1}^{N}q_{ik}\left [ \left ( \frac{t_{ji}-\mu _{jk}}{\sigma_{jk}  }\right )^{2} -1 \right ]
\end{eqnarray}

where $q_{ik}$ is the soft assignment weight of feature $t_{i}$ to the $k^{th}$ Gaussian and defined as
\begin{eqnarray}
q_{ik}=\frac{exp\left [-\frac{1}{2}\left ( t_{i} -\mu_{k}\right )^{T}\Sigma _{k}^{-1}\left ( t_{i} -\mu_{k}\right ) \right ]}{\sum_{t=1}^{K}exp\left [-\frac{1}{2}\left ( t_{i} -\mu_{t}\right )^{T}\Sigma _{t}^{-1}\left ( t_{i} -\mu_{t}\right ) \right ]}
\end{eqnarray}

Just as the image representation, the dimension of Fisher vector is $2KD$, $K$ is the number of components in the  GMM, and $D$ is the dimension of local feature descriptor. After ${l_2}$ normalization on Fisher vector, the Euclidean distance is calculated to find similarity between two images. 

%

\subsubsection{Sparse coding on deep features}
\label{sec:matrix_dictionary}

The compressed (sparse) sensing framework allows us to learn a set of over-complete bases $D$  and sparse weights $\phi$ such that the feature matrix $T$ can be represented by a linear combination of these basis vectors:

\begin{eqnarray}
\mathop {\arg \min }\limits_{D,\phi } \frac{1}{n}\sum\limits_{i = 1}^n {\left( {\frac{1}{2}\left\| {T - D \cdot {\phi _i}} \right\|_F^2{\text{ }} + \lambda {\text{ }}{{\left\| {{\phi _i}} \right\|}_1}} \right)} 
\end{eqnarray}

The k-SVD algorithm \cite{Jiang2011} comprises of two stages -- first, a sparse coding stage (either using matching pursuit or basis pursuit) and second, a dictionary update stage. In the first stage when the dictionary $D$ is fixed, we solve for $\phi$ using orthogonal matching pursuit (OMP). Briefly, OMP recovers the support of the weights $\phi$ using an iterative greedy algorithm that selects at each step the column of $D$ that is most correlated with the current residual. Practically, we initialise the residual ${r_k}$, subsequently computing the column that reduces the residual the most ${j^ * }$ and finally adding this column to the support $ {I_k}$

\begin{eqnarray}
  {r_k} & = & T - D{\phi _{k - 1}} \nonumber \\
  {j^ * } & = & \mathop {\arg \min }\limits_{\mathop {j = 1 \ldots n}\limits_\phi  } {\left\| {{r_k} - {d_j}\phi } \right\|_2} \nonumber \\ 
  {I_k} & = & {I_{k - 1}} \cup {j^ * } 
\end{eqnarray}

Iterating through these equations for a pre-specified number of iteration, we can update the sparse weight $\phi$. After obtaining the sparse weights, we use a dictionary update stage where we update only one column of $D$ each time. The update for the $k$-th column is,

\begin{eqnarray}
\left\| {T - D \cdot \phi } \right\|_F^2 & = & \left\| {T - \sum\limits_{j = 1}^K {{D_j} \cdot \phi _j^T} } \right\|_F^2 \nonumber \\
 & = &  \left\| {\underbrace {\left( {T - \sum\limits_{j \ne k}^{} {{D_j} \cdot \phi _j^T} } \right)}_{{E_k}} - {D_k} \cdot \phi _k^T} \right\|_F^2
\end{eqnarray}

In order to minimize $\left\| {{E_k} - {D_k} \cdot \phi _k^T} \right\|_F^2$ we decompose (SVD) ${E_k}$ as $UW{V^T}$. Using this decomposition we utilize a rank-1 approximation to form ${{ d}_k} = {u_0}$ and ${\phi _k} = {w_0}{v_0}$. We can then iterate this for every column of $D$. Sparsity is attained by collapsing each ${\phi _k}$ only to non-zero entries, and constraining ${E_k}$ to the corresponding columns.

For image retrieval, each local deep feature can be encoded by sparse weights $\phi$.  Such an image signature can be represented by max pooling of a set of $\phi$, followed by measuring a distance between such sets. 

\subsubsection{Tensor factorization of deep features}
\label{sec:tensor_factorization}

In the earlier section, we relied on an alternate minimization of the dictionary and the loadings (weights) to yield a convex optimization problem. Specifically, the last convolutional layer was used to form a dictionary without reference to the tensorial (multi-linear) representation of feature spaces obtained from the deep convolutional network. Thus, our goal is to approximate these tensors as a sparse conic combinations of atoms that have been learnt from a dictionary comprising the entire image training set. In other words, we would like to obtain an over-complete dictionary such that each feature tensor can be represented as a weighted sum of a small subset of the atoms (loadings) of the dictionary.

There are two decompositions that are used for factorizing tensors, one is based on Tucker decomposition whilst the second is known as  Canonical Decomposition/Canonical Polyadic Decomposition (CANDECOMP/CPD), also known as Parallel Factor Analysis (PARAFAC). In the former, tensors are represented by sparse core tensors with block structures. Specifically, $\mathcal{T}$ is approximated as a multi-linear transformation of a small ``core'' tensor $\mathcal{G}$ by factor matrices $A$ \cite{Kolda2009},

\begin{eqnarray}
\mathcal{T} = \mathcal{G}{ \bullet _1}{A^{(1)}}{ \bullet _2}{A^{(2)}}{ \ldots _N}{A^{(N)}} \triangleq \left[\kern-0.15em\left[ {\mathcal{G};{A^{(1)}},{A^{(1)}} \ldots {A^{(N)}}} 
 \right]\kern-0.15em\right]
 \label{eqn:tucker}
\end{eqnarray}

In the latter, a tensor $\mathcal{T}$ is written as a sum of $R$ rank-1 tensors, each of which can be written as the outer product of $N$ factor variables i.e.,

\begin{eqnarray}
\mathcal{T} = \sum\limits_{r = 1}^R {a_r^{(1)}}  \otimes a_r^{(2)} \ldots a_r^{(N)} \triangleq \left[\kern-0.15em\left[ {{A^{(1)}},{A^{(2)}} \ldots {A^{(N)}}} 
 \right]\kern-0.15em\right]
\end{eqnarray}
 
It is canonical when the rank $R$ is minimal; such a decomposition is unique under certain conditions \cite{Domanov2013}. Even then, due to numerical errors, factorization of a feature matrix obtained using a deep neural network results in a non-unique factorization. Thus, CPD proves inadequate for unique feature encoding. We therefore utilize a factorization that is similar to a 2D-PCA albeit lifted for multi-linear objects. Specifically, we use t-SVD \cite{Kilmer2013} to factorize the feature matrices. 

\subsubsection*{Based on t-SVD}
\label{sec:tsvd}

The t-product between two tensors, $\mathcal{T}_1$ and  $\mathcal{T}_2$,

\begin{eqnarray}
  {\mathcal{T}_1} * {\mathcal{T}_2} & \equiv & {\text{fold}}({\text{circ}}({\mathcal{T}_1}) \cdot {\text{unfold}}({\mathcal{T}_2})) \nonumber \\
  \mathcal{T} & = & \mathcal{U} * \mathcal{S} * \mathcal{V}^T \hfill  
\end{eqnarray}

where, ${\text{circ}}( \cdot )$ creates a block circulant matrix and the unfold operator matricizes the tensor on the tube (3rd) axis. $\mathcal{S}$ is a f-diagonal tensor that contains the eigen-tupules of the covariance tensor on its diagonal whereas, the columns of $\mathcal{U}$ are the eigenmatrices of the covariance tensor. The images in the training set are inserted as the second index of the tensor. In other words, using t-SVD, we obtain an orthogonal factor dictionary of the entire training set. Ultimately, a projection of the mean-removed input tensor (i.e., a single feature tensor, ${\mathcal{T}_{{\text{test}}}}$)  on the orthogonal projector (${U^T} * {\mathcal{T}_{{\text{test}}}}$) forms the tensor encoding of each image. The Frobenius norm then measures the closeness between a pair of images (individual tensor projections). Computation efficiency is guaranteed since the comparison between the test image and the database is measured in Fourier domain -- the t-product utilizes a fast Fourier transform algorithm in its core \cite{Nvidia2007}.

Another feature encoding that we consider is the partition of each individual tensor i.e.,  $\mathcal{T} =  \mathcal{L} + \mathcal{P}$ where, $\mathcal{L}$ is a low-rank tensor and $\mathcal{P}$ is a sparse tensor. We have $\mathcal{L} = \mathcal{U}_{1:r} * \mathcal{S}_{1:r} * \mathcal{V}^T_{1:r}$ and $\mathcal{P} = \mathcal{T} - \mathcal{L}$.  $r$ denotes the truncation index of the singular components.

\subsubsection*{Based on mPCA}
\label{sec:mpca}

For high-order tensor analysis, multilinear Principal Component Analysis(mPCA) or High Order Singular Value Decomposition(HOSVD) \cite{Vasilescu2003,Lu2008} compute a set of orthonormal matrices associated with each mode of a tensor -- this is analogous to the orthonormal row and column space of a matrix computed by the matrix PCA/SVD. 

In a Tucker decomposition (Eqn. \ref{eqn:tucker}) if the factor matrices are constrained to be orthogonal the input tensor can be decomposed as a linear combination of rank-1 tensors.  Given a set of N-order training tensor $\mathcal{T}$, the objective of mPCA is to find N linear projection matrices that maximize the total tensor scatter (variance) in the projection subspace. When factor matrices ${A^{(N)}}$ in Eqn. \ref{eqn:tucker} are orthogonal then $\left\| \mathcal{T} \right\|_F^2 = \left\| \mathcal{G} \right\|_F^2$. Each query (test) tensor $\mathcal{T}_q$ can then be projected using $\mathcal{Y} = {\mathcal{T}_q}{ \times _1}{{\mathbf{A}}^{(1)T}}{ \times _2}{{\mathbf{A}}^{(2)T}} \ldots { \times _N}{{\mathbf{A}}^{(N)T}}$ where the bold-faced matrices represent a low-dimensional space. The objective then becomes learning a set of matrices $\mathbf{A}$ to admit

\begin{eqnarray}
\underset{\mathbf{A}^{(1)}...\mathbf{A}^{(N)}}{\arg\max}\sum_{m=1}^{M}\left \| \mathcal{Y}_{m,train} - \overline{\mathcal{Y}}\right \|_{F}^{2}
\end{eqnarray}

$\overline{\mathcal{Y}}$ is the mean tensor. Since the projection to an $N^{th}$ order tensor subspace consists of N projections to N vector subspaces, in Ref. \cite{Lu2008},  optimization of the N projection matrices can be solved by finding each $A^{(n)}$ that maximizes the scatter in the n-mode vector subspace with all other $A$-matrices fixed. The local optimization procedure can be iterated until convergence.  

\section*{Experiments}
\label{sec:experiments}

In this section, five deep feature encoding methods are evaluated for image retrieval on the DTD dataset. Images are resized to same size (256 x 256), deep feature is extracted from last convolutional layer of a pre-trained VGG-M. For Fisher vector and sparse coding, deep features are flattened as a set of 1D local features. For t-SVD, deep features are represented as 2D feature maps, and treated as [HxW,1,D] data structures (see Methods). After encoding and ${l_2}$ normalization, the Euclidean distance is calculated to find similarity between two images. 

To evaluate image retrieval, Mean Average Precision (MAP) on top 10 rankings are calculated. Two images per category i.e., a total of 94 images are selected as queries from the test dataset. The dataset retrieved includes 3760 images from DTD training and validation datasets. MAP on DTD is listed in Table \ref{table:dtdmAP}.  An example of the retrieval obtained with each method is shown in Figure 2. On each case 10 images are displayed. Top left image is the query used. The rest of images are arranged by similarity to query image as obtained with each encoding method.

\begin{figure}
 \centering \includegraphics[width=0.5\textwidth, height=1.5in]
 {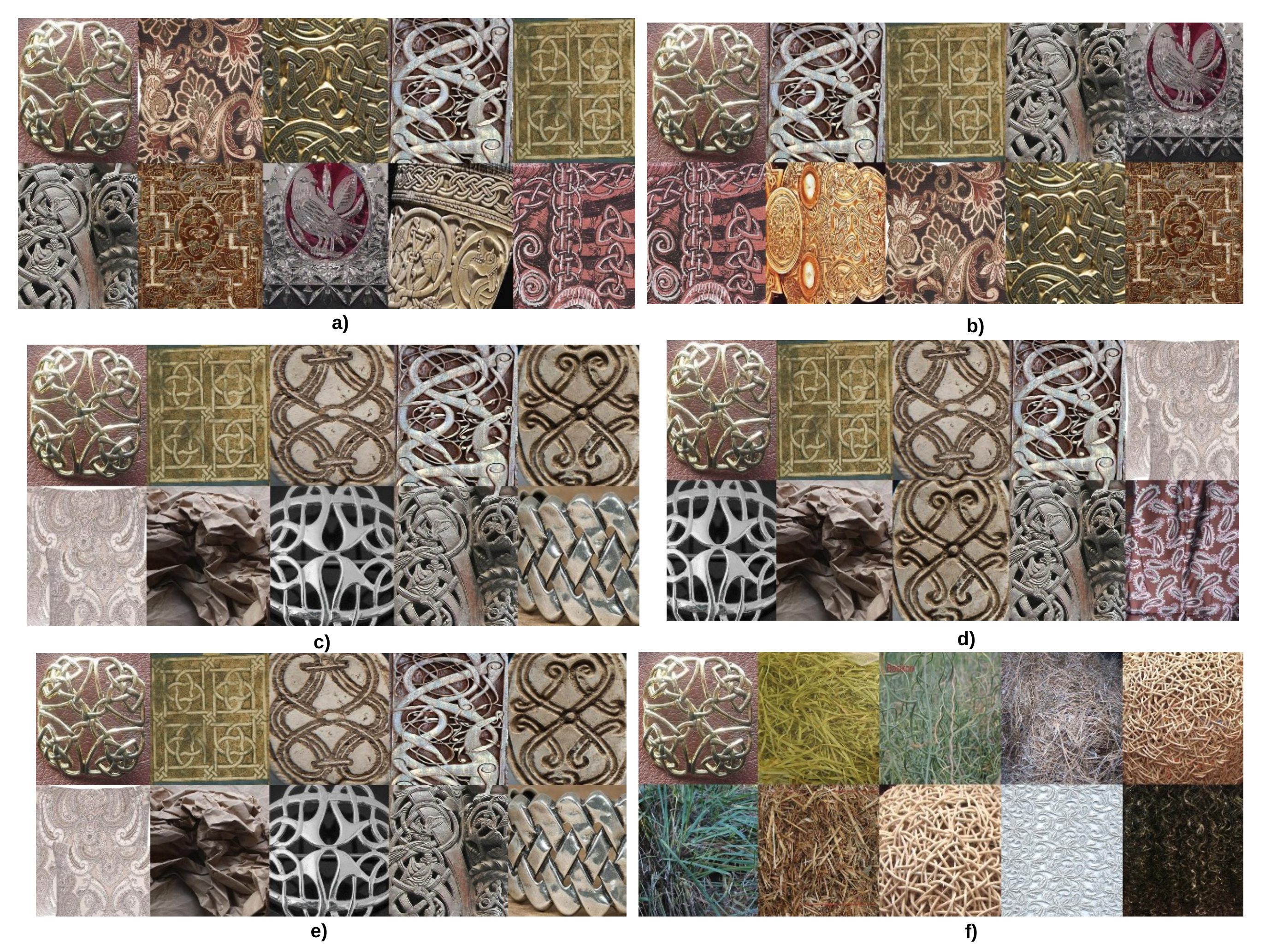}
 \caption{Retrieved results on DTD: a) Fisher vector  b) Sparse coding c) t-SVD d) mPCA  e) Low rank tensor f) Raw tensor}
 \label{fig_Retrieval}
\end{figure}

Image retrieval amounts to searching large databases to return ranked list of images that have similar properties to that of the query image. Here, we reason that in contrast to raw feature tensors (i.e., without any encoding of the feature maps), their encodings either using Fisher vector, sparse dictionaries or multi-linear tensors increases the average precision of retrieval. Table \ref{table:dtdmAP}  shows that multi-linear encodings based on t-SVD, mPCA or low-rank decomposition of individual images all have similar fidelity whilst performing very close to information geometric encoding of the feature vectors. 

Sparse coding supersedes other methods in terms of average precision. This is because the dictionary learnt using k-SVD is matched to the underlying image distribution (i.e., by learning the sparse weights) whereas the tensor dictionaries (t-SVD or mPCA) use orthogonal matrices as dictionaries without the added constraint to finesse the weights, or to add additional structure in terms or sparsity, low-rank or non-negativity.

As shown in Table \ref{table:dtdmAP}, computing mPCA tensor encodings are computationally efficient in contrast to sparse dictionaries or learning a probabilistic model for Fisher vector encodings. Combined with reasonable retrieval performance, tensor encodings of deep neural features make them a contender for commercial infrastructures.      

The code was implemented in Matlab 2015a under linux with Intel Xeon CPU E5-2640 @ 2.00GHz and 125G RAM. Sandia's Tensor toolbox \cite{Bader2015}, KU Leuven's Tensorlab \cite{Vervliet2016} and TFOCS \cite{Becker2011} were used to implement our algorithms.

\begin{table}[]
\centering
\caption{Average precision for the DTD dataset. Raw tensors are feature tensors without any encoding.}
\label{table:dtdmAP}
\begin{tabular}{lllll}
\textbf{Methods}                & \textbf{top-1} & \textbf{top-5} & \textbf{top-10} & \textbf{time taken}  \\
Fisher Vector          & 0.56  & 0.52  & 0.48 & 12ms  \\
Sparse Coding       & 0.62  & 0.49  & 0.44   & 35ms \\
t-SVD dictionary       &  0.53     & 0.42      & 0.38   & 2188ms    \\
mPCA dictionary       &  0.53      & 0.43       & 0.39    & 5ms    \\
Low Rank       &  0.51     & 0.42      & 0.38   & 967ms    \\
Raw Tensor              & 0.41      & 0.35       & 0.31      
\end{tabular}
\end{table}

\section*{Conclusion}
\label{sec:conclusion}

Feature encoding is crucial for comparing images (or videos, text, etc.) that are similar to one another. Our experiments show that whilst sparse encoding of a feature tensor proves to be the most efficient encoding for retrieval, having no encoding grossly reduces the average precision. Taking the multi-linear properties of the feature tensor  improves retrieval, and the fidelity is at par with Fisher encoding. We further show that computing such multi-linear representation of the feature tensor is computationally much efficient than constructing a sparse dictionary or learning a probabilistic model.  

The sparse dictionary encoding has the highest average precision due to the added flexibility of learning the weights as well as imposing the structural constraint of sparsity. Fisher vector encoding has the second highest precision because of its ability to capture the information geometry of the underlying probabilistic model. Specifically, the Fisher tensor encodes the underlying Riemannian metric which augments the encoding with the curvature of the underlying distribution. The multi-linear approaches based on t-SVD, mPCA and low-rank decomposition perform at par with Fisher vectors as they encode the third and higher order interaction in the feature tensors. Comparison of the compute time tells us that amongst all of the methods, encoding deep feature tensors using mPCA is the most time-efficient algorithm.

Our results have not exploited the geometry exhibited by the tensors, for example, one can calculate lagged covariance tensors from the feature tensor -- these tensors exhibit a Riemann geometry due to their positive definite structure. Therefore building a dictionary of co-variance tensors using t-SVD, wherein an Augmented Lagrangian alternating direction method can be employed to learn a sparse representation, is the next viable step to our work. We anticipate that such a multi-linear overcomplete dictionary should at the very least have increased precision to that of the Fisher encoding method. In so far, the last convolutional layer was used to form a dictionary without reference to the earlier layers of the deep neural network. In fact a step forward would be to constrain the construction of an image signature with tensorial information from an earlier layer. The tensor methods rely on factorizing large tensors, especially those that emerge from deep neural networks. Yet, GPU implementation of the Fourier transform in the form of \textit{cuFFT} enables us to build a scalable commercial solution (\url{https://www.cortexica.com/technology/}).

\bibliographystyle{ACM-Reference-Format}
\bibliography{cortexica_fisher}

\begin{appendices}
\section{Tensor Norms}
\label{appendix:tnorm}

Let $t_{ijk}$ be the elements of tensor $\mathcal{T}$, then the Frobenius norm is ${\left\| \mathcal{T} \right\|_F} = {\left\| {vec(\mathcal{T})} \right\|_2} = \sqrt {\sum\limits_i {\sum\limits_j {\sum\limits_k {t_{ijk}^2} } } }$. The nuclear (trace) norm is defined as ${\left\| \mathcal{T} \right\|_*} = trace(\sqrt {{\mathcal{T}^T}\mathcal{T}} ) = \sum\limits_{i = 1}^{\min \{ m,n\} } {{\sigma _i}}$.  $\sigma$ are the singular values of $\mathcal{T}$.

\section{Cortexica's findSimilar Technology}

\begin{figure}
 \centering \includegraphics[width=0.5\textwidth, height=1.5in] {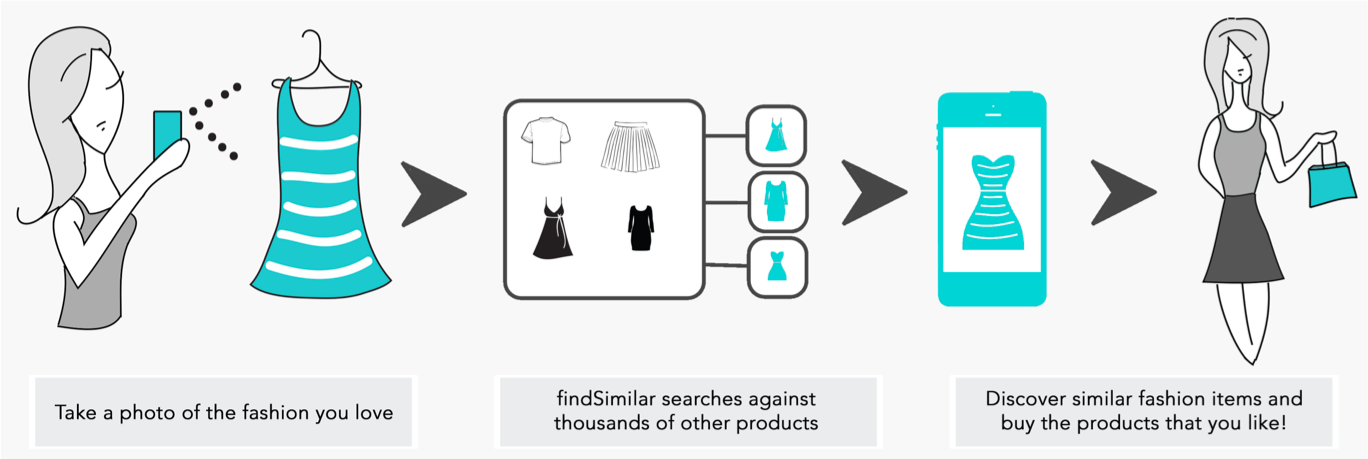}
 \centering \includegraphics[width=0.5\textwidth, height=1.5in]
 {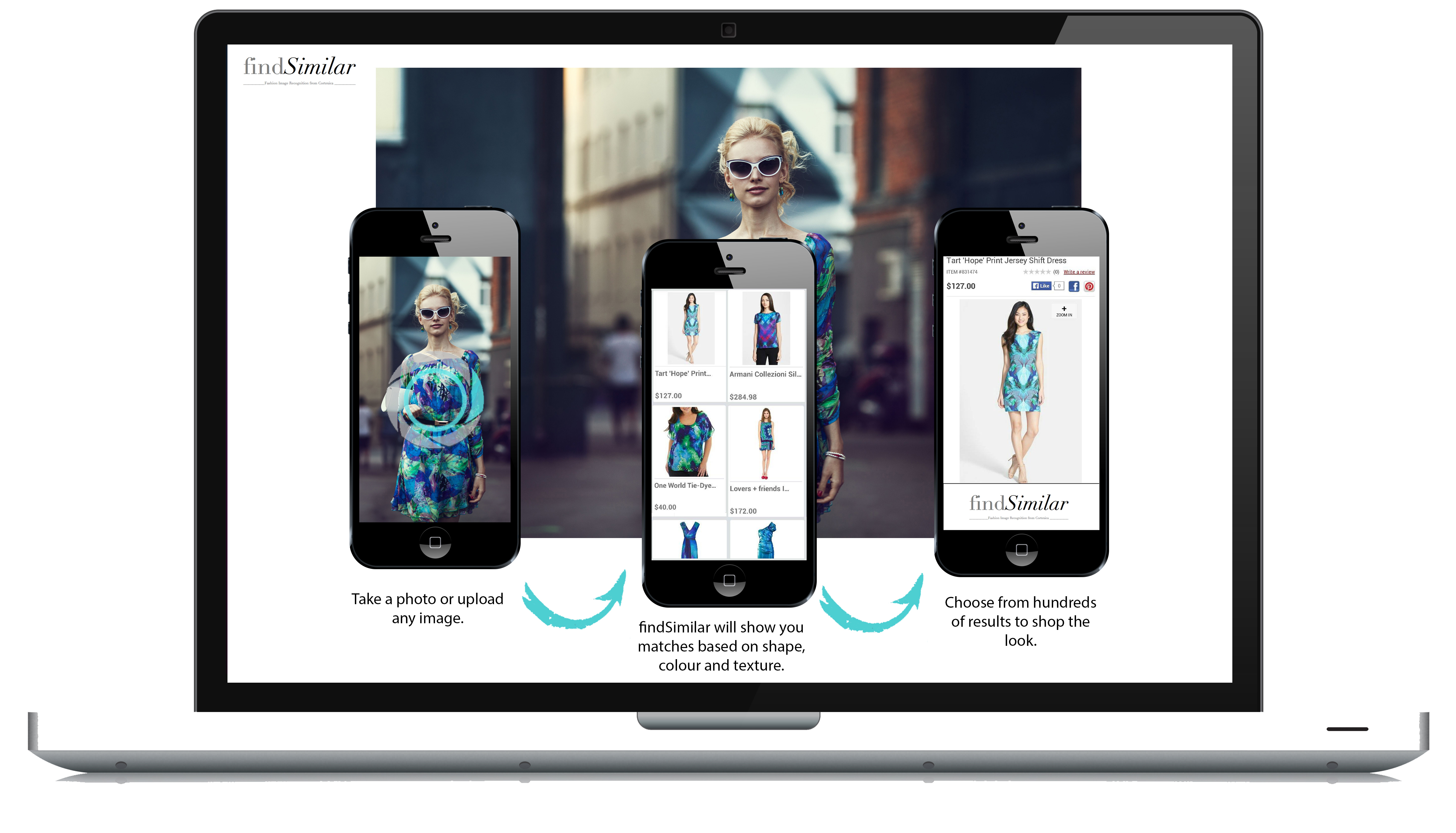}
 \caption{The findSimilar technology: A consumer takes a photograph of a clothing item. Using proprietary versions of feature encodings that leverage deep-learning as well as multi-scale analysis, the retailer presents similar items from the database.}
 \label{fig:fsimilar}
\end{figure}

\begin{figure}
  \begin{subfigure}[b]{0.3\textwidth}
   \includegraphics[width=\textwidth]{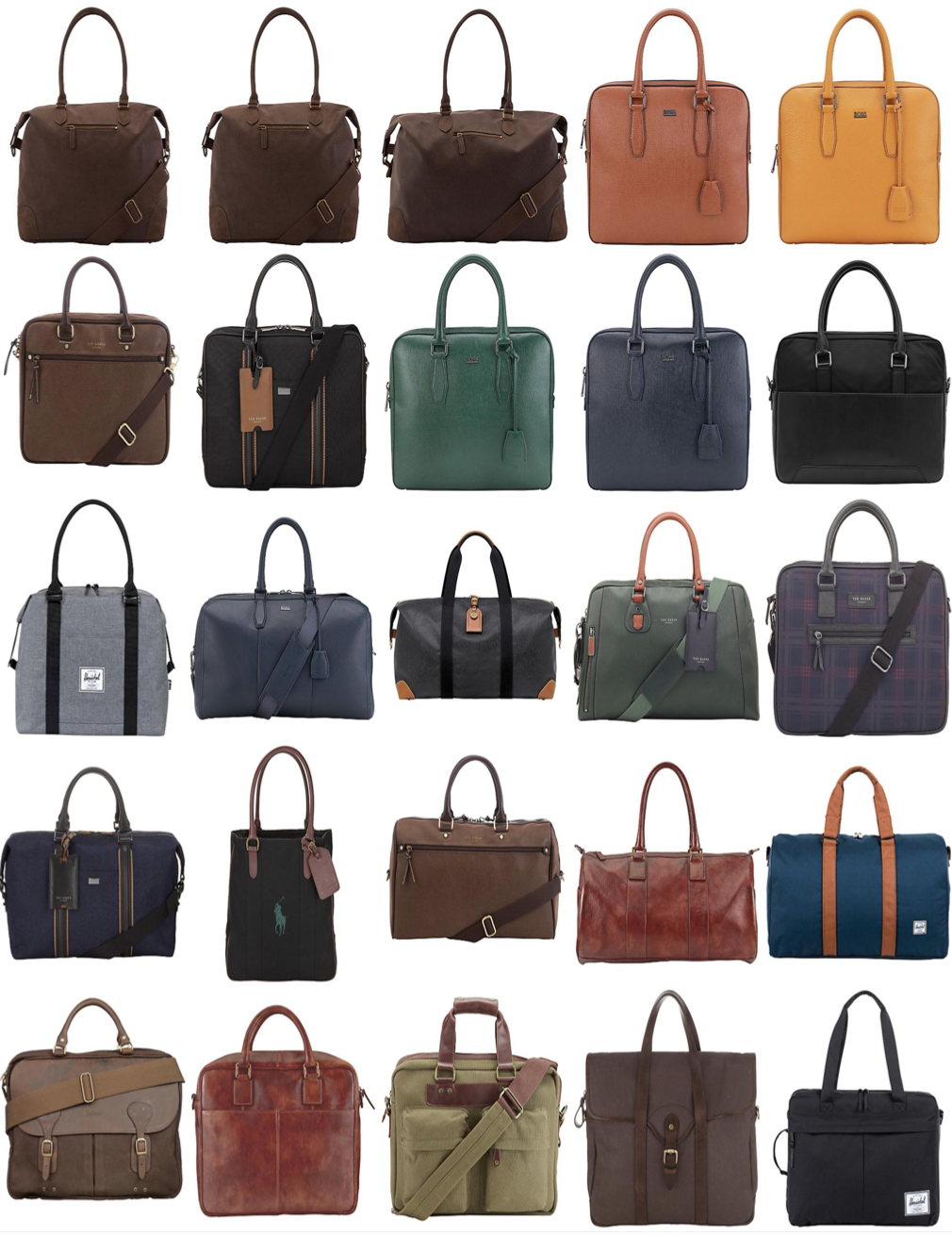}
   \caption{Retrieval of similar bags}
   \label{fig:bags}
  \end{subfigure}
  \begin{subfigure}[b]{0.3\textwidth}
   \includegraphics[width=\textwidth]{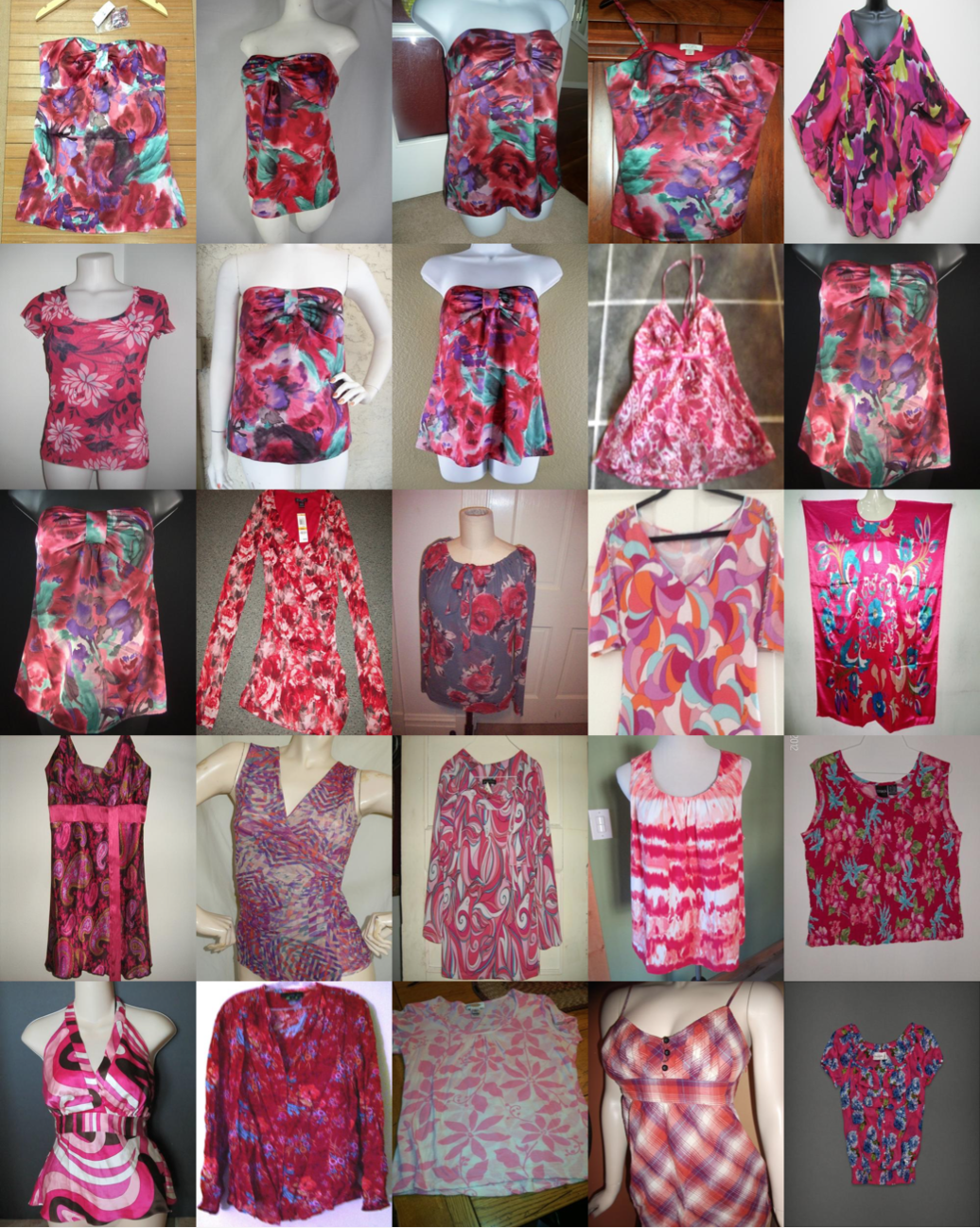}
    \caption{Retrieval of similar dresses}
   \label{fig:dress}
  \end{subfigure}
 \caption{Feature encodings: Each image is encoded using a (proprietary) combination of encodings described in this paper, along with other patented descriptors. Shown here are examples wherein the query is the top-left item and a ranked list is returned based on similarity. }
 \label{fig:simexamps}
\end{figure}

\end{appendices}

\end{document}